\documentclass[aps,prd,twocolumn,10pt,eqsecnum,showpacs,nofootinbib,floatfix,superscriptaddress,longbibliography]{revtex4-1}

\usepackage{lmodern}
\usepackage[T1]{fontenc}
\usepackage[utf8]{inputenc}
\usepackage{amsmath,amssymb,amsfonts}
\usepackage{mathrsfs}
\usepackage[colorlinks]{hyperref}
\usepackage[all]{hypcap}
\usepackage{calc}
\usepackage{xspace}
\usepackage{verbatim}
\usepackage{tikz}
\usepackage{microtype}
\usepackage{enumerate}

\usetikzlibrary{arrows}
\usetikzlibrary{decorations.markings}

\usepackage{environ}
\makeatletter
\newsavebox{\measure@tikzpicture}
\NewEnviron{scaletikzpicturetowidth}[1]{%
  \def\tikz@width{#1}%
  \def\tikzscale{1}\begin{lrbox}{\measure@tikzpicture}%
  \BODY
  \end{lrbox}%
  \pgfmathparse{#1/\wd\measure@tikzpicture}%
  \edef\tikzscale{\pgfmathresult}%
  \BODY
}
\makeatother

\newcommand{\TM}{T\mathcal{M}}
\newcommand{\LtTM}{\Lambda^2\TM}
\newcommand{\TsM}{T^*\!\mathcal{M}}

\newcommand{\cd}{\nabla}
\newcommand{\scri}{\mathscr{I}^+}
\newcommand{\dualKY}{\ensuremath{{}^*\!f}}

\newcommand{\tlin}{\texorpdfstring{\\}{ }}

\newcommand{\figholonomy}{
\begin{figure}[tb]
\begin{center}
\begin{scaletikzpicturetowidth}{\columnwidth}
\begin{tikzpicture}
[
y=1pt, x=1pt, yscale=-\tikzscale, xscale=\tikzscale,
fiberstyle/.style={draw=blue, line width=2pt},
basemstyle/.style={draw=black, loosely dotted, line width=1pt,
  fill=gray!25!white},
basepathstyle/.style={draw=black, line width=2pt,
  decoration={markings, mark= at position 0.5 with {\arrow{stealth}},},
  postaction={decorate}},
fiberpathstyle/.style={draw=red, dashed, line width=2pt,
  decoration={markings, mark= at position 0.5 with {\arrow{stealth}},},
  postaction={decorate}}
]
  \path[basemstyle]
    (27.9299,918.3484) .. controls (99.0432,949.5653) and (162.6508,879.4139) ..
    (233.9246,887.9080)  node [below] {patch of $\mathcal{M}$}
    .. controls (291.5807,894.4958) and (364.5307,855.7964) ..
    (357.4063,790.3532) .. controls (344.2588,721.3047) and (256.6888,699.1025) ..
    (207.3656,724.6811) .. controls (146.9998,741.2882) and (95.0975,790.9648) ..
    (35.7306,810.4849) .. controls (1.9208,824.9305) and (-4.4902,891.5319) ..
    (27.9299,918.3484) -- cycle;

  \coordinate (P) at (138.2567,888.9221);
  \coordinate (Q) at (314.4927,817.0052);
  \coordinate (R) at (246.5272,755.3621);
  \coordinate (S) at (73.4524,826.4887);

  \coordinate (overP1) at (142.9777,827.4454);
  \coordinate (overP2) at (141.8673,777.3240);
  \coordinate (overQ1) at (319.2344,752.2009);
  \coordinate (overR1) at (250.4787,694.5093);
  \coordinate (overS1) at (77.4039,726.9114);

  \coordinate (overP) at (128.8179,685.4303);
  \coordinate (overQ) at (305.0091,687.0015);
  \coordinate (overR) at (236.6367,639.9748);
  \coordinate (overS) at (54.7092,665.3505);

  \node[below left] at (P) {$\mathcal{P}$};
  \node[below right] at (Q) {$Q$};
  \node[above left] at (R) {$R$};
  \node[below left] at (S) {$S$};

  \node[above] at (overP) {\textcolor{blue}{$\pi^{-1}(\mathcal{P})$}};

  \path[basepathstyle]
    (P) .. controls (169.8685,879.0434) and (279.3522,824.0686) .. (Q)
    node [midway, above] {$u^{a}$};
  \path[basepathstyle]
    (Q) .. controls (296.9396,796.7983) and (263.5472,777.5374) .. (R);
  \path[basepathstyle]
    (R) .. controls (212.9110,775.8627) and (119.8766,808.2299) .. (S);
  \path[basepathstyle]
    (S) .. controls (102.8966,847.1185) and (121.6604,860.8666) .. (P)
    node [midway, below left] {$-v^{b}$};

  \path[fiberpathstyle]
    (overP1) .. controls (176.3057,810.0424) and (278.0707,765.2790) .. (overQ1);
  \path[fiberpathstyle]
    (overQ1) .. controls (283.2806,720.6816) and (265.9524,707.4270) .. (overR1);
  \path[fiberpathstyle]
    (overR1) .. controls (217.6814,701.7092) and (121.0603,714.0590) .. (overS1);
  \path[fiberpathstyle]
    (overS1) .. controls (96.8945,735.2933) and (123.9849,753.8534) ..  (overP2);

  \path[fiberstyle]
    (P) .. controls (153.1681,789.4624)
    and (128.8179,685.4303) .. (overP);
  \path[fiberstyle]
    (Q) .. controls (329.5083,753.3863)
    and (305.0091,687.0015) .. (overQ);
  \path[fiberstyle]
    (R) .. controls (261.1752,698.6700)
    and (236.6367,639.9748) .. (overR);
  \path[fiberstyle]
    (S) .. controls (98.1654,747.4438) and
    (54.7092,665.3505) .. (overS);

\end{tikzpicture}
\end{scaletikzpicturetowidth}
\end{center}
  \caption{(Color online)
    Transport about a coordinate rectangle $\mathcal{P}QRS$ in a small
    patch of the base manifold $\mathcal{M}$.  The coordinate
    rectangle is defined by the vector $u^{a}$, which points from
    $\mathcal{P}$ to $Q$, and $v^{b}$, which points from $\mathcal{P}$
    to $S$.  Over each point $p\in\mathcal{M}$ is a 10-dimensional
    fiber $\pi^{-1}(p)$ (depicted in blue) in the bundle
    $\mathcal{B}$.  Transport around $\mathcal{P}QRS$ on the base
    manifold (depicted in black) with the new connection $\tilde{\cd}$
    gives transport through $\mathcal{B}$ (depicted in red, dashed).
    This  induces a linear transformation acting on
    $\pi^{-1}(\mathcal{P})$, as expressed in
    Eq.~\eqref{eq:delta-X-expr}.
  }
  \label{fig:holonomy}
\end{figure}
}

\begin{document}

\author{\'Eanna \'E.~Flanagan}
\affiliation{Cornell Center for Astrophysics and
Planetary Science (CCAPS), Cornell University, Ithaca, New York 14853, USA}

\author{David A.~Nichols}
\email{david.nichols@cornell.edu}
\affiliation{Cornell Center for Astrophysics and
Planetary Science (CCAPS), Cornell University, Ithaca, New York 14853, USA}

\author{Leo C.~Stein}
\affiliation{Theoretical Astrophysics,
Walter Burke Institute for Theoretical Physics,
California Institute of Technology, Pasadena, California
91125, USA}

\author{Justin Vines}
\affiliation{Max-Planck-Institut f\"ur Gravitationsphysik,
Albert-Einstein-Institut, Am M\"uhlenberg 1, 14476 Golm, Germany}

\title{Prescriptions for measuring and transporting\tlin{}local angular momenta in general relativity}

\begin{abstract}

For observers in curved spacetimes, elements of the dual space of the set of linearized Poincar\'e transformations
from an observer's tangent space to itself can be naturally interpreted as local linear and angular momenta.
We present an operational procedure by which observers can measure
such quantities using only information
about the spacetime curvature at their location.
When applied by observers near spacelike or null infinity
in stationary, vacuum, asymptotically flat spacetimes, there is a sense in
which the procedure yields
the well-defined linear and angular momenta of the spacetime.

We also describe a general method by which observers can transport
local linear and angular momenta from one point to another,
which improves previous prescriptions.
This transport is not path independent in general, but becomes path
independent for the measured momenta in the same limiting regime.
The transport prescription is defined in terms of differential equations,
but it can also be interpreted as parallel transport in a particular
direct-sum vector bundle.
Using the curvature of the connection on this bundle,
we compute and discuss the holonomy of the transport law.
We anticipate that these measurement and transport definitions
may ultimately prove useful for clarifying the physical interpretation of the
Bondi-Metzner-Sachs charges of asymptotically flat spacetimes.

\end{abstract}

\maketitle

\section{Introduction}

Asymptotically flat spacetimes in general
relativity have an infinite-dimensional group of
asymptotic symmetries, rather than the ten translations,
rotations, and boosts of flat Minkowski spacetime
(see~\cite{Geroch1977} for a review of asymptotically flat
spacetimes).
This larger symmetry group, the Bondi-Metzner-Sachs (BMS)
group~\cite{Bondi1962, Sachs1962, Sachs1962b}, differs from the
Poincar\'e group of flat spacetime because the BMS group contains an
infinite family of ``angle-dependent translations'' called the
supertranslations, rather than the four spacetime translations of the
Poincar\'e group.
The quotient of the BMS group by the supertranslations is isomorphic to the Lorentz group,
 just as the quotient of the Poincar\'e group by the
translations is the Lorentz group.

Associated with each BMS symmetry generator ${\vec \xi}$ is a corresponding Noether-like charge $Q({\vec \xi})$,
which is not conserved, but whose change between different
times (or more precisely, cuts of null infinity) is determined by a
flux formula~\cite{Ashtekar1981, Dray1984, Wald2000}.
These charges can be computed in terms of integrals of the spacetime curvature
over cuts of null infinity.  They can in principle be measured by families
of observers near null infinity who measure the spacetime geometry in their
vicinities and who communicate their results to one another so as to evaluate
the charge integrals.
On the other hand, observers who attempt to measure Poincar\'e-covariant asymptotic charges
will in general recover the BMS charges associated with some observer-dependent Poincar\'e subgroup of the BMS group.
The purpose of this paper is to explore in more detail how such measurements
can be made and to understand their observer dependence.
One motivation for this exploration is to try to understand more deeply the physical interpretation of the BMS charges themselves.

Consider an observer at an event $\mathcal{P}\in\mathcal{M}$ in the
spacetime manifold $\mathcal{M}$.
The mathematical space of a linear and angular momentum as measured by that
observer is $\mathcal{G}_{\mathcal{P}}^*$, the space dual to
linearized Poincar\'e transformations (affine transformations) from
the tangent space $T_{\mathcal{P}}\mathcal{M}$ to itself.  This space
can be naturally parameterized in terms of pairs of tensors $(P^a,J^{ab})$ at
${\cal P}$, with $J^{(ab)}=0$, which represent the linear and angular momentum 
about the observer's location~\cite{Flanagan2014}.
How can such local linear and angular momenta be defined and measured?

One approach to such definitions is the following:
Suppose that a Poincar\'e subgroup of the BMS group has been specified;
for example, it could
be the subgroup associated with a stationary region of future null infinity
($\scri$).  Let $\mathfrak{g}$ be the corresponding algebra of
generators ${\vec \xi}$, where $\mathfrak{g} \subset \mathfrak{bms}$
is a subalgebra of the BMS algebra, and $\mathfrak{g} \simeq
\mathfrak{iso}(3,1)$ is isomorphic to the Poincar\'e algebra.
Suppose also that one has a prescription for extending BMS generators ${\vec \xi}$ (which are vector fields defined on $\scri$) into the interior of the spacetime.
An example of such a prescription associated with the retarded Bondi coordinate conditions is given in Ref.~\cite{Barnich:2010eb}; many other prescriptions exist.  We now define tensors $P^a$, $J^{ab}$ at ${\cal P}$ by
\begin{equation}
Q({\vec \xi}) = P^a \xi_a({\cal P}) + \frac{1}{2} J^{ab} \nabla_{[a} \xi_{b]}({\cal P})
\label{eq:bmsdef}
\end{equation}
for any ${\vec \xi}$ in $\mathfrak{g}$.
Here the left-hand side is the BMS charge, which is a linear function of 
${\vec \xi}$.
On the right-hand side, we can identify the Poincar\'e algebra
$\mathfrak{g}$ with the
space of values of $\xi_a$ and $\nabla_{[a} \xi_{b]}$ at ${\cal P}$, and 
thereby determine the coefficients $P^a$ and $J^{ab}$.
The definition~\eqref{eq:bmsdef} clearly depends on the choice of prescription for extending generators into the interior of the spacetime,
but one would expect the leading-order terms in an expansion of the 
prescription in powers of $1/r$, as $r\to \infty$,
would be independent of this choice.

A different approach to defining local linear and angular momenta at a point ${\cal P}$
was explored by two of the authors in Ref.~\cite{Flanagan2014} 
(henceforth Paper~I) and will be extended and refined in this paper.
They defined a procedure by which an observer could
measure quantities $(P^a,J^{ab})$ from the spacetime geometry at her location.
By contrast, the definitions of $P^a$ and $J^{ab}$ in terms of BMS
charges are nonlocal functionals of the spacetime geometry.
The procedure was designed
to recover the correct momentum and angular momentum of a linearized, vacuum,
asymptotically flat stationary spacetime when used near
future null infinity, up to corrections of order
$M/r$, where $M$ is the mass of the spacetime, and $r$ is the distance
to the source as measured by the asymptotic observer.

Paper~I also defined a prescription for transporting elements of 
${\cal G}_{\cal P}^*$ along paths in spacetime.
This definition was based on a rule for transporting vectors along curves by a 
generalization of parallel transport which was called ``affine transport.''
The angular-momentum transport law can also be defined explicitly in the 
following way, as shown in Appendix~A of Paper~I:
given a curve with tangent $k^a$, the pair $(P^a, J^{ab})$ is transported 
along the curve using the differential equations\footnote{%
Equation~\eqref{eq:TransportOld} corrects a sign error in Appendix~A of the published version of Paper~I.}
\begin{subequations}
\begin{align}
k^a \nabla_a P^b = {} & 0 \, ,\\
k^a \nabla_a J^{bc} = {} & 2 P^{[b} k^{c]} \, .
\end{align}
\label{eq:TransportOld}%
\end{subequations}
This transport law allows two observers at different spacetime locations
to compare values of angular momentum that they measure, albeit in a curve-dependent fashion.
In addition, changes with time of angular momentum can be compared by
transporting the angular momentum about a closed curve in spacetime composed
of the two observers' worldlines and
two spacelike curves connecting their locations.
This process amounts to computing a holonomy of the affine transport
law.  In Paper~I, it was shown that such holonomies contain two
parts: the normal holonomy associated with parallel transport of tensors
(denoted ${\Lambda^a}_b$), and an inhomogeneous part $\Delta \xi^a$ related to a
``displacement vector'' that depends upon the curve.
This displacement vector satisfies $k^a \nabla_a \Delta \xi^b=k^a$
with initial condition $\Delta \xi^a =0$.
The momentum and angular momentum transform under this holonomy by the
following laws:
\begin{subequations}
\begin{align}
P^a \rightarrow {} & {\Lambda^a}_b P^b \, ,\\
J^{ab} \rightarrow {} & {\Lambda^a}_c {\Lambda^b}_d (J^{cd}
- 2 \Delta\xi^{[c} P^{d]}  ) \, .
\end{align}
\label{eq:PJtransform}%
\end{subequations}
When the map ${\Lambda^a}_b$ reduces to the identity and $\Delta\xi^a$
is zero (to leading order in $M/r$, for example), then the asymptotic
observers agree on the momentum and angular momentum of the spacetime.
However, a nontrivial ${\Lambda^a}_b$ or $\Delta\xi^a$ indicates that
spacetime curvature produces an obstacle to observers arriving at a
consistent definition of a linear and angular momentum of the spacetime.

One of the primary goals of Paper~I was to use the local measurement
procedure for angular momentum and the holonomy of the affine
transport equation to understand the physics behind what is often
called the ``supertranslation ambiguity'' of angular momentum in
general relativity.
The ambiguity refers to the fact that while there is a four-parameter
translation subgroup of the supertranslations, there is, in general,
no preferred
Poincar\'e subgroup of the BMS group.
As a result, the charges in general relativity associated with the
six-parameter factor group of the BMS group depend, in general, on a
smooth function on the 2-sphere rather than a four-parameter origin.
Stationary spacetimes are an exception in this regard:
they possess a preferred Poincar\'e subgroup of the BMS group
with associated Poincar\'e charges.

When the measurement and transport procedures are applied to
``sandwich-wave'' spacetimes (in which a burst of linearized
gravitational waves of finite duration with memory pass through
Minkowski space),
the results of the measurements are observer dependent.
Furthermore, Paper~I showed that this observer dependence is related
to the supertranslation that relates the shear-free cuts in the
Minkowski space before the burst to those after the burst (which is,
in essence, just the memory effect; see, e.g., \cite{Strominger:2014pwa}).
More specifically, the generalized holonomy contains a nontrivial
inhomogeneous part which is a function of the
aforementioned supertranslation evaluated at the observers' locations
and of the separation of the observers.
In this context, the measurement procedure gives the linear momentum
of the spacetime and an observer-dependent angular momentum of the
spacetime that depends on a four-parameter choice of origin of the
spacetime.
The holonomy gave---in the form of a Poincar\'e
transformation---information about the BMS supertranslation (at the
location of the two observers), which creates an obstruction to defining
a consistent notion of angular momentum that depends upon a
four-parameter origin.

There are two closely related limitations of the measurement and
transport procedure of Paper~I outside of the context of the
sandwich wave spacetimes described above.
First, in the measurement procedure, the angular momentum is only defined to
a fractional accuracy of order $M/r$ which, because the angular momentum about
the point scales as $M r$, implies that there are errors in the angular
momentum of order $M^2$.
These errors, however, are of the same size as the observer dependence
arising from the memory effect that is of physical interest.
Second, a nontrivial holonomy does not necessarily imply the existence of
ambiguities related to BMS transformations.
For example, for certain spacelike closed curves in the asymptotic region of a
Schwarzschild black hole, the generalized holonomy is nontrivial, even though
stationary spacetimes have a preferred Poincar\'e subgroup of the BMS group
(and hence a well-defined angular momentum).

In this paper, we refine the definitions of Paper~I of both the local
measurement of angular momentum and the transport procedure, in order that the
measurement be sufficiently accurate to capture supertranslation/memory effects,
and in order that nontrivial asymptotic holonomies only arise because
of BMS-type observer dependence.
The refined definitions are sufficiently accurate that in
vacuum, stationary, asymptotically flat
spacetimes, observers near $\scri$ will agree
upon their measured linear and angular momentum
(this includes the location of the source's center of mass, which is now
measured with an accuracy $\sim M^2/r$).

The paper is organized as follows.
In Sec.~\ref{sec:Transport}, we define the new transport equation for
tensors $P^a$ and $J^{ab}$, and we describe specific path-independent
solutions of this transport equation.
In Sec.~\ref{sec:Momentum}, we define a prescription for measuring local
linear and angular momentum from spacetime curvature and show
that it gives the expected answer in appropriate limiting regimes.
In Sec.~\ref{sec:Bundle}, we describe how our transport equation
can be understood as a connection on a certain bundle, and we
compute the curvature of this bundle (and therefore, also the
holonomy of an infinitesimal loop).
We conclude in Sec.~\ref{sec:Conclusions}.
We use throughout geometric units ($G=c=1$) and the conventions of
Wald~\cite{Wald1984}.

\section{Transport equations for angular momentum}
\label{sec:Transport}

As a generalization of~\eqref{eq:TransportOld}, we will
consider transport equations along curves with tangent $k^a$ of the
form
\begin{subequations}
\begin{align}
k^a \nabla_a P^b ={} & -\kappa R^{b}{}_{acd} J^{cd} k^a \, ,\\
k^a \nabla_a J^{bc} ={} & 2 P^{[b} k^{c]} \, ,
\end{align}
\label{eq:TransportNew}%
\end{subequations}
where $\kappa$ is a real constant.
From the point of view of the theory of differential equations, these
transport equations have nice properties.
For all values of $\kappa$, these equations are linear in $P^a$ and
$J^{ab}$ and reparameterization invariant under changes in the tangent
$k^a$.
Solutions of the equations yield a linear map between the spaces of tensors $(P^a,J^{ab})$ at the initial point and at the final point of the curve.  When $\kappa=0$, this linear map reduces to the standard
action of a Poincar\'e transformation on $(P^a,J^{ab}$), as discussed in
Paper~I.
For nonzero values of $\kappa$, however, the
corresponding map has a more involved form.

Transport equations of the form~\eqref{eq:TransportNew} arise in
several different contexts in
the field of general
relativity:
\begin{enumerate}[(i)]

\item The case $\kappa=0$ is the simplest version of an angular momentum transport law
consistent with the properties of angular momentum in special
relativity.  Its properties were studied in Paper~I.

\item The $\kappa=1/2$ transport equations have the same form as the Mathisson-Papapetrou
equations~\cite{Mathisson1937, Papapetrou1951}, when $P^a$ and
$J^{ab}$ are taken to be the linear and angular momentum of a
particle (rather than of the spacetime) and the curve is the
particle's worldline.

\item The $\kappa=1/2$ transport equations
are also dual to the Killing transport equations, in the following sense:
Suppose that $A_a$ and $B_{ab} = B_{[ab]}$ are tensors which satisfy the Killing transport equations along the curve (as would be
the case if there were a Killing vector field $\xi^a$ on the spacetime and $A_a$ and $B_{ab}$ were defined by
$A_a = \xi_a$ and $B_{ab} =\nabla_a\xi_b= \nabla_{[a} \xi_{b]}$).
In addition, suppose that $P^a$ and $J^{ab}$ satisfy the transport 
equations~\eqref{eq:TransportNew} with $\kappa =1/2$.  
Then the generalized momentum $ P^a A_a +  J^{ab} B_{ab}/2 $
is conserved along the curve~\cite{Harte2008}.

\item In this paper, we will use the $\kappa = -1/4$ transport equations
  to define a prescription for transporting angular momentum.  We will
  also show that observers who use this prescription will arrive at a
  mutually consistent definition of angular momentum, near future null
  infinity in stationary, vacuum, asymptotically flat spacetimes.

\end{enumerate}
While most of this paper focuses on the case $\kappa=-1/4$, our calculations in Sec.~\ref{sec:Bundle} below are valid for all
values of $\kappa$, and might prove useful in some of these other contexts.

In this paper, we will be most interested in situations where the transport
equations admit solutions that are independent of the path used to
transport the tensors $P^a$ and $J^{ab}$ throughout the
asymptotically flat region of a stationary, vacuum spacetime.
If they admit such path-independent solutions, then there will be a linear and angular
momentum of the spacetime that different observers can measure and
transport consistently.
A sufficient condition for such curve-independent solutions to exist
is if there are solutions to the partial differential equations
\begin{subequations}
\begin{align}
\nabla_a P^b ={} & {}-\kappa {R^b}_{acd} J^{cd} \, ,\\
\nabla_a J^{bc} ={} & {}2 P^{[b} {\delta^{c]}}_a
\end{align}
\label{eq:FieldEqs}%
\end{subequations}
throughout the region of interest.

In the next two subsections, we will show that solutions to the
equations~\eqref{eq:FieldEqs} do exist.
First, in Sec.~\ref{subsec:global}, we show that there is an exact solution in the Kerr spacetime
when $\kappa=-1/4$, which is defined throughout the entire spacetime.
Next, in Sec.~\ref{subsec:Stationary}
we show that approximate solutions exist in general,
asymptotically flat, stationary spacetimes near future null
infinity (or equivalently, in this case, spacelike infinity), again
only when $\kappa=-1/4$.  The existence of those approximate solutions
is sufficient to allow asymptotically consistent measurements of
angular momentum, as we discuss in more detail in Sec.~\ref{subsec:PrescripProps} below.

\subsection{Global solution in the Kerr spacetime}
\label{subsec:global}

The global solution to~\eqref{eq:FieldEqs} in the Kerr spacetime is
a consequence of the relationships between the Killing-Yano (KY) tensor
and the timelike Killing field that this spacetime admits.
Many of the properties that we use here are well known, and can be
found in Floyd's thesis~\cite{Floyd1974} or in more recent review
papers~\cite{Yasui2011}.
We begin by noting that a second-rank KY tensor is an antisymmetric
tensor $f_{ab}$ that satisfies the differential equation
\begin{equation}
\nabla_{(a} f_{b)c} = 0 \, .
\label{eq:KYdef}
\end{equation}
As a consequence of~\eqref{eq:KYdef}, a KY tensor also satisfies the
integrability condition
\begin{equation}
\nabla_a \nabla_b f_{cd} = -\frac{3}{2} {R^e}_{a[bc} f_{d]e} \, .
\label{eq:IntegCond}
\end{equation}
The dual of the KY tensor will be denoted by
\begin{equation}
\dualKY^{ab} = \frac{1}{2}\epsilon^{abcd} f_{cd} \, .
\label{eq:dualKY}
\end{equation}
In the Kerr spacetime, the divergence of the dual of the KY
tensor is related to the timelike Killing field by
\begin{equation}
\xi^b = \frac{1}{3} \nabla_a \dualKY^{ab} \, .
\label{eq:KYvector}
\end{equation}
Using Eqs.~\eqref{eq:KYdef}--\eqref{eq:KYvector},
we can show after some calculation that the gradients of the fields
$\xi^a$ and $\dualKY^{ab}$ satisfy the set of equations
\begin{subequations}
\begin{align}
\nabla_a \xi^b &=  -\frac{1}{4} R^{b}{}_{acd} \dualKY^{cd} \, , \\
\nabla_a \dualKY^{bc} &= -2 \xi^{[b} {\delta^{c]}}_a \, .
\end{align}
\label{eq:KYFieldEqs}%
\end{subequations}

We immediately see that the identification
$(P^{a},J^{ab})=(\xi^{a},-\dualKY^{ab})$ exactly solves the transport
equations~\eqref{eq:FieldEqs} with $\kappa = -1/4$.  However, this
exact solution does not have the physical interpretation we seek.  In
the limit $r \to \infty$, the tensor $-\dualKY^{ab}$ has two pieces, one which
acts like an intrinsic angular momentum, and one like an orbital
angular momentum about the spacetime point.
The relative sign of these two pieces is the opposite of what it should
be for $-\dualKY^{ab}$ to be the asymptotic angular momentum,
as noted by Floyd~\cite{Floyd1974}.
Thus, this exact solution is not directly relevant for our purposes.

\subsection{Asymptotic approximate solutions in stationary asymptotically
flat spacetimes}
\label{subsec:Stationary}

We now show that arbitrary stationary, asymptotically flat spacetimes
admit approximate asymptotic solutions to the partial differential
equations~\eqref{eq:FieldEqs} with $\kappa = -1/4$.

We adopt Bondi coordinates $(u,r,\theta^A)$,
in which the coordinate $u$ foliates $\scri$ by
shear-free cuts, $r$ is an affine parameter along null rays, and
$\theta^A$ are arbitrary coordinates on the unit 2-sphere.
We specialize to a center-of-momentum Bondi coordinate system.
It follows (see, e.g., \cite{Flanagan:2015pxa}) that the spacetime
metric can be written in the form
\begin{align}
ds^2 = & -\left(1-\frac{2M}{r}-\frac{2\mathcal M}{r^2}\right)du^2 - 2dudr
\nonumber \\
& {} + r^2 h_{AB} d\theta^A d\theta^B + \frac{4}{3}N_A d\theta^A du
+ \ldots\, .
\label{eq:BondiMetricStat}
\end{align}
Here $M$ is a constant, the ellipsis denotes higher-order terms in a series in $r^{-1}$,
$h_{AB}$ is a metric on the unit 2-sphere, and $D_A$ will denote a covariant
derivative on the 2-sphere.  Also $N_A(\theta^A)$ is a function
satisfying $(D_B D^B + 1) N_A = 0$
(i.e., it is composed of $\ell=1$ spherical harmonics), and $\mathcal M$ 
satisfies $6 \mathcal M = -D^A N_A$, which follows from Einstein's
equations.

Next, we make the following two ansatzes for the form of the solution.
First we assume that the Lie derivative of $P^a$ and $J^{ab}$ with
respect to $\partial_u$ vanishes.  This is a natural requirement since
$\partial_u$ is a Killing vector.  Second, we assume the following
large-$r$ expansions of the solutions:
\begin{subequations}
\begin{align}
P^{\hat \mu} ={} & P^{\hat \mu}_{(0)}(\theta^A) + \frac Mr P^{\hat \mu}_{(1)}(\theta^A)
+ O \left( \frac{M^2}{r^2} \right) ,\\
J^{{\hat \mu}{\hat \nu}} ={} & \frac{r}{M} J^{{\hat \mu}{\hat \nu}}_{(0)}(\theta^A) +
J^{{\hat \mu}{\hat \nu}}_{(1)}(\theta^A) + O\left( \frac{M}{r} \right) ,
\end{align}
\label{eq:PJseries}%
\end{subequations}
where the hatted indices refer to components of the tensors on the basis
${\vec e}_{\hat \mu}$ given by
\begin{equation}
{\vec e}_{\hat u} = \partial_u, \ \ \ \
{\vec e}_{\hat r} = \partial_r, \ \ \ \
{\vec e}_{\hat A} = \frac{1}{r} \partial_A.
\end{equation}

Now substituting the ansatz \eqref{eq:PJseries}
into the differential equations \eqref{eq:FieldEqs} and matching order
by order in
powers of $1/r$, we immediately find several
constraints on $P^\mu_{(0)}$, $J^{\mu\nu}_{(0)}$,
$P^\mu_{(1)}$, and $J^{\mu\nu}_{(1)}$.
These constraints are that $\kappa = -1/4$, $P^\mu_{(1)} = 0$,
$J^{ur}_{(0)} = M P^u_{(0)}$, $P^r_{(0)} = 0$, $J^{AB}_{(0)} = 0$, and
$J^{rA}_{(0)} = -J^{uA}_{(0)} = 0$.
With these conditions imposed, the equations further simplify, and it is
then easy to show that $P^A_{(0)} = 0$, that $\partial_\mu P^u_{(0)} = 0$,
and $\nabla_\rho^{(0)} J^{\mu\nu}_{(1)} = 0$, where $\nabla_\rho^{(0)}$ is
the covariant derivative operator of Minkowski spacetime in the coordinates
$(u,r,\theta^A)$.
It therefore follows that $P^u_{(0)}$ is a constant (which need not
coincide with the Bondi mass $M$ of the spacetime though).
Because $J^{\mu\nu}_{(1)}$ is antisymmetric and satisfies the same equation 
as a covariantly constant tensor in Minkowski spacetime, it can be
parameterized by six constants.

We have shown, therefore, that for the specific value $\kappa=-1/4$,
there exist asymptotic solutions to the equations~\eqref{eq:FieldEqs} in a
stationary spacetime and that their expansion in Bondi coordinates has
the form
\begin{subequations}
\begin{align}
P^\mu \partial_\mu ={} & P^u_{(0)} \partial_u + \ldots \, , \\
J^{\mu\nu} \partial_\mu \otimes \partial_\nu ={} &
r P^u_{(0)} \left[ \partial_u \otimes \partial_r - \partial_r \otimes \partial_u \right]
\nonumber \\
& +  J^{\mu\nu}_{(1)} \partial_\mu \otimes \partial_\nu + \ldots \, ,
\end{align}
\label{eq:StationaryPJ}%
\end{subequations}
where $P^u_{(0)}$ is a constant and $J^{\mu\nu}_{(1)}$ is parameterized
by six constants.
Thus, there is in fact a seven-parameter family of solutions\footnote{%
The
family of solutions contains seven parameters rather than the ten parameters
associated with the Poincar\'e group, because the
equations~\eqref{eq:FieldEqs} require the solution $P^a$ to be
asymptotically
proportional to the Bondi 4-momentum of the spacetime, eliminating the
boost freedom.}
$(P^a,J^{ab})$ that can be transported by Eq.~\eqref{eq:TransportNew}
with $\kappa=-1/4$ in an asymptotically path-independent manner in the region of the
spacetime described by the metric~\eqref{eq:BondiMetricStat}.

In the next section, we will discuss a prescription for how observers
can measure quantities ($P^a$, $J^{ab}$) at their locations, and we will argue 
that the existence of the approximate solutions~\eqref{eq:StationaryPJ} can be
used to demonstrate consistency of such measurements made by different 
observers.

\section{Procedure for measuring angular momentum}
\label{sec:Momentum}

In this section, we define
a prescription for how an observer at an
event $\mathcal P$ can measure an element of
$\mathcal{G}_{\mathcal{P}}^*$, the space dual to linearized Poincar\'e
transformations on the tangent space $T_{\mathcal{P}}\mathcal{M}$.
That element can be parameterized as a pair of tensors $(P^a, J^{ab})$
at $\mathcal P$, as discussed in Paper~I, which can be interpreted as
approximate versions of the linear and angular momentum of the
spacetime about the observer's location.
The prescription requires several assumptions about the geometry near
$\mathcal P$ and, consequently, is applicable only in
certain situations.  The definition of the prescription is given in
in Sec.~\ref{subsec:MomPrescrip}, and
some of its properties are discussed in
Sec.~\ref{subsec:PrescripProps}.

\subsection{Prescription for measuring angular momentum}
\label{subsec:MomPrescrip}

The steps of the prescription are as follows:

\begin{enumerate}[(i)]

\item\label{enum:Step1}
Measure all the components of the Riemann tensor $R_{abcd}$ and
of its gradient $\nabla_a R_{bcde}$ at the event $\mathcal P$.

\item\label{enum:Step2} Compute the curvature invariants
\begin{subequations}
\begin{align}
K_1 \equiv{}& R_{abcd} R^{abcd} , \\
\mathcal K_1 \equiv{}& \nabla_a R_{bcde} \nabla^a R^{bcde} \, ,
\end{align}
\label{eq:invariants}%
\end{subequations}
which we assume to satisfy $K_1 >0$ and $\mathcal K_1>0$.
Then compute quantities $M$ and $r$ using (cf.~Footnote~8 of Paper~I)
\begin{subequations}
\begin{align}
M ={} & \frac{15\sqrt 5 (K_1)^2}{4\mathcal K_1^{3/2}} \left(1 -
\frac{15\sqrt 3 K_1^{3/2}}{4\mathcal K_1}\right) \, ,\\
r ={} & \sqrt{\frac{15 K_1}{\mathcal K_1}} \left(1 -
\frac{5\sqrt 3 K_1^{3/2}}{4\mathcal K_1}\right) \, .
\label{eq:rdef}
\end{align}
\label{eq:mrdef}%
\end{subequations}

\item Repeat steps (\ref{enum:Step1}) and (\ref{enum:Step2}) at nearby
spacetime points, so as to measure the gradient $\nabla_a r$ of the
quantity $r$.

\item
Assuming that the vector $\nabla_a r$ is spacelike,
define the unit vector $n^a$ in the direction of $\nabla_a r$ by
$n^a = (N_1)^{-1} \nabla^a r$, where
$N_1 = \sqrt{(\nabla^a r) (\nabla_a r)}$.
Next, compute the quantity
\begin{equation}
y^a = - (r + M) n^a \, ,
\label{eq:zdef}
\end{equation}
which the observer interprets as a perpendicular displacement vector
from her location to the position of the center-of-mass worldline of
the source.

\item Construct the symmetric tensor $H_{ab}$ from
\begin{equation}
H_{ab} = R_{acbd} n^c n^d \, .
\label{eq:Hdef}
\end{equation}
Compute the eigenvectors $\zeta^a$ and eigenvalues $\lambda$ of $H_{ab}$
from $H_{ab} \zeta^b = \lambda \zeta_a$.
From the definition~\eqref{eq:Hdef}, one of the eigendirections will be
$\zeta^a = n^a$ with corresponding eigenvalue $\lambda=0$.
We will assume that at least one eigenvector has a strictly positive
eigenvalue, and we denote the eigendirection corresponding to the largest
eigenvalue by $t^a$.
It follows that this vector is orthogonal to $n_a$ (i.e., $t^a n_a =0$).

\item Assuming that the vector $t^a$ is timelike, next define a
unit, future-directed timelike vector $v^a$ by $v^a = (N_2)^{-1} t^a$.
The normalization $(N_2)^{-1}$ is defined from $(N_2)^2 = -t_a t^a$ and
the sign of $N_2$ is chosen so that $v^a$ is future directed.

\item
Compute the magnetic part of the Weyl tensor along $v^a$
\begin{equation}
B_{ae} = -\frac{1}{2} \epsilon_{abcd} {C^{cd}}_{ef} v^b v^f \, .
\end{equation}
From this construct a spin vector $S^a$ by
\begin{equation}
S^a = \frac{r^4}{2} {B^a}_b n^b - \frac{2r^4}{3} (B_{bc}n^b n^c) n^a \, ,
\label{eq:Sadef}
\end{equation}
and a 4-velocity vector $u^a$ by
\begin{equation}
u^a = \left(1+\frac Mr\right)v^a + \frac 1{Mr} {\epsilon^a}_{bcd} v^b S^c n^d
\, .
\label{udef}
\end{equation}

\item Define the angular momentum and linear momentum to be
\begin{subequations}
\begin{align}
P^a ={} & M u^a \, ,\\
J^{ab} ={} & {\epsilon^{ab}}_{cd} u^c S^d + 2 y^{[a} P^{b]} \, .
\label{eq:Jdef}
\end{align}
\end{subequations}
Finally from $(P^a, J^{ab})$ compute an element of
$\mathcal G_{\mathcal P}^*$ using the definition~(2.1) of Paper~I
specialized to ${\vec x}_0=0$.

\end{enumerate}

\subsection{Motivation for and properties of the prescription}
\label{subsec:PrescripProps}

We now discuss the motivations for the measurement prescription
and some of its properties.

First, we note that the prescription refines the prescription given in
Paper~I, at subleading order in $M/r$, in a number of ways.
First, the expressions~\eqref{eq:mrdef} for $M$ and $r$
contain higher-order correction terms
  constructed from the curvature invariants~\eqref{eq:invariants}.
Second, the spin in~\eqref{eq:Sadef} is constructed using the magnetic part of
the Weyl tensor rather than the symmetric tensor $H_{ab}$ of
Eq.~\eqref{eq:Hdef} and a pseudoscalar curvature invariant.
Finally, the four-velocity of the source is given by Eq.~\eqref{udef} rather than
being proportional to $t^a$.

We next discuss some of the motivations for these refinements.
Consider the following three properties of algorithms to produce tensor fields $P^a$ and $J^{ab}$ from the local
spacetime geometry:

\begin{enumerate}[(i)]

\item\label{enum:p1}
In the context of linearized gravity, the algorithm reproduces the expected 
answers for stationary, vacuum spacetimes near future null infinity, in the 
limit $r \to \infty$.

\item\label{enum:p2}
The specification of the algorithm does not require any preferred lengthscale or
a choice of spacetime orientation.

\item\label{enum:p3}
Consider the tensor fields $P^a$ and $J^{ab}$ obtained by applying the algorithm
to a stationary, vacuum region of an asymptotically flat spacetime near future null infinity.
When these tensor fields are expanded in powers of $1/r$, the leading and subleading terms yield a solution
of the transport equations~\eqref{eq:FieldEqs} with $\kappa = -1/4$ to
the accuracy discussed in Sec.~\ref{subsec:Stationary}.  In other
words, they yield a specific element of the seven parameter
family~\eqref{eq:StationaryPJ} of approximate asymptotic solutions.

\end{enumerate}

The prescription of Paper~I satisfies properties (\ref{enum:p1}) and (\ref{enum:p2}),
while the refined algorithm of this paper is designed to additionally
satisfy property (\ref{enum:p3}).  This requirement necessitates improving the
accuracy of the algorithm, from leading order in $M/r$ to subleading
order in $M/r$.
More precisely, when applied to an arbitrary vacuum,
stationary, asymptotically flat spacetime,
the definitions~\eqref{eq:mrdef}
of $M$ and $r$ yield respectively the Bondi mass and the radial
coordinate of the Bondi system~\eqref{eq:BondiMetricStat}
specialized to the center-of-mass frame condition $D_A N^A=0$, up to
fractional
errors of order $\sim M^2/r^2$ and $\sim S^2 M^{-2} r^{-2}$ in both
quantities.
In particular, when applied to the Kerr spacetime,
the algorithm reproduces the ADM mass and the Boyer-Lindquist radial coordinate to an equivalent accuracy.
The 4-momentum $P^a$ and angular momentum $J^{ab}$ of the new algorithm have fractional errors\footnote{%
Here when we refer to the fractional error in the tensor fields we do not mean to imply
the existence of ``correct'' versions to which our prescription can be compared.
Instead, we mean that the leading and subleading pieces of $P^a$ and $J^{ab}$ 
have been carefully specified, but higher-order pieces have not.}
of the same order.
This implies that components of $P^a$ and $J^{ab}$ in an orthonormal basis have errors
that scale as $\sim M^3/r^2$ and $\sim M^3/r$, respectively, as compared to the errors $\sim M^2/r$
and $\sim M^2$ in Paper~I.  Orthonormal-basis components of the displacement vector $y^a$ to the center of mass
have errors of order $\sim M^2/r$, rather than the $\sim M$ errors of Paper~I.

Consider now the tensor fields $P^a$ and $J^{ab}$ produced by the
algorithm in the stationary, vacuum region of an asymptotically flat spacetime near
future null infinity.  Since property (\ref{enum:p3}) is satisfied,
when $(P^a, J^{ab})$ are transported
by the transport equations~\eqref{eq:TransportNew} with $\kappa=-1/4$, these tensors
will be transported in a path-independent way to the above
accuracy---the same accuracy with which they are measured.
It therefore follows that observers will find consistency between their
measured values of linear and angular momentum in the limit $r \to \infty$.

Next, we discuss the extent to which the measurement algorithm is unique.
As discussed in Paper~I, imposing the requirements (\ref{enum:p1}) and
(\ref{enum:p2}) does not determine a unique algorithm in linearized
gravity, since the information about the asymptotic charges is encoded
redundantly in the values of the Riemann tensor and its first two
derivatives at a point.  Nevertheless, in Paper~I, the leading-order pieces of
$P^a$ and $J^{ab}$ were uniquely determined by the requirement
(\ref{enum:p1}).

Similarly, here, imposing the requirements (\ref{enum:p1}),
(\ref{enum:p2}), and (\ref{enum:p3}) does not yield a unique algorithm,
because of redundancy in how information is encoded in the Riemann
tensor and its gradients at a point.\footnote{%
For example, one could
have chosen the version of the expansion~\eqref{eq:rdef} appropriate
for the radial coordinate $r'=\sqrt{x^2+y^2+z^2}$ of harmonic, quasi-Cartesian,
Cook-Scheel coordinates~\cite{CookScheel1997}
instead of the Boyer-Lindquist radial coordinate.
This would require
subleading modifications to the prescription for measuring
the vectors $y^a$, $S^a$, and $u^a$ in order to satisfy property
(\ref{enum:p3}).}
Nevertheless,
the leading and subleading pieces of $P^a$ and $J^{ab}$ are uniquely
determined.  In other words, all algorithms which satisfy the three
properties yield the same solution out of the seven-parameter family
of approximate solutions discussed in Sec.~\ref{subsec:Stationary}.
This is because the seven free parameters in the
solutions~\eqref{eq:StationaryPJ} are fixed by imposing that the
algorithm satisfy the requirements (\ref{enum:p1}) and
(\ref{enum:p2}).  We note that if the requirement (\ref{enum:p2})
is dropped, ambiguities in the algorithm do arise at subleading order,
of the form
$y^a \to y^a +\lambda \epsilon^{abcd} u_b J_{cd}/M$, where $\lambda$
is a positive dimensionless constant and $y^a \equiv - J^{ab} P_b/M^2$.

\section{Transport equations as parallel transport in a fiber bundle}
\label{sec:Bundle}

We now turn to studying the transport equation~\eqref{eq:TransportNew} from a different point of view.
A mathematically equivalent but conceptually different perspective on
the transport equation can come from considering a direct-sum (Whitney-sum) 
bundle over the manifold $\mathcal{M}$.
Specifically we will consider the bundle
$\mathcal{B} \equiv \TM\oplus \LtTM$, of which
the pair $(P^{a},J^{[ab]})$ is a section.
We will take derivatives of
sections of this bundle, which will require us to discuss a
number of connections on $\mathcal B$ and other bundles.

The Levi-Civita connection $\cd$ extends from $\TM$ and $\TsM$ to all
tensor-product bundles via the usual definition~\cite{Wald1984}.
We will continue to use the same symbol $\cd$
for all of the distinct connections acting on different bundles.  We
can also extend $\cd$ to a connection on $\mathcal{B}$ in the simplest
possible way, by letting $\cd$ act diagonally on each summand in
$\mathcal{B}$.

Now recall that the space of connections is an affine space; thus, if
$\cd$ is a connection, then so is $\tilde{\cd}=\cd+\Gamma$, where
$\Gamma$ is a one-form taking values in the space of linear maps.
Starting from the transport equations~\eqref{eq:TransportNew}, we can
define a new connection on $\mathcal{B}$ via
\begin{equation}
  \label{eq:cd-tilde-column-vec}
  \tilde{\cd}_{a}
  \begin{pmatrix}
    P^{b} \\ J^{bc}
  \end{pmatrix}
  \equiv
  \cd_{a}
  \begin{pmatrix}
    P^{b} \\ J^{bc}
  \end{pmatrix}
  +
  \begin{pmatrix}
    \kappa {R^b}_{acd} J^{cd} \\ - 2 P^{[b} {\delta^{c]}}_a
  \end{pmatrix}\,.
\end{equation}
Because the second term on the right is linear in the vector
$(P^{a},J^{ab})$, there is clearly a corresponding linear-map-valued
one-form, $\Gamma$.
Let us introduce an index notation for tensors in the different
bundles: specifically, we will use lower-case Latin
indices for $\TM$ and $\TsM$, and capital Latin indices for
$\mathcal{B}$ and $\mathcal{B}^{*}$.  A capital index will thus range
over the collection of lower-case indices, e.g.~$B=(b,[bc])$.
In this notation, we may denote the pair as
\begin{equation}
  X^{B} \equiv
  \begin{pmatrix}
    P^{b} \\ J^{bc}
  \end{pmatrix}
  \begin{matrix}
    \mbox{\tiny{B=b}} \\ \mbox{\tiny{B=[bc]}}
  \end{matrix} \, ,
\end{equation}
and we can rewrite Eq.~\eqref{eq:cd-tilde-column-vec} in the form
\begin{equation}
  \tilde{\cd}_{a} X^{B} = \cd_{a} X^{B} + \Gamma_{a}{}^{B}{}_{D} X^{D}\,.
\end{equation}
Now we would like to read off the form of $\Gamma_{a}{}^{B}{}_{D}$
from Eq.~\eqref{eq:TransportNew}.
By inserting a Kronecker delta and permuting indices, we find
\begin{equation}
\Gamma_{a}{}^{B}{}_{D} =
\begin{matrix}
  \mbox{\tiny{B=b}} \\ \mbox{\tiny{B=[bc]}}
\end{matrix}
\overset{
\begin{matrix}
  \makebox[\widthof{$2\delta_{a\vphantom{d}}^{[b} \delta^{c]}_{d}$}]{\tiny{D=d}}
& \makebox[\widthof{$\kappa {R^b}_{ade}$}]{\tiny{D=[de]}}
\end{matrix}
}{
\begin{pmatrix}
0 & \kappa R^{b}{}_{ade} \\
2\delta_{a\vphantom{d}}^{[b} \delta^{c]}_{d} & 0
\end{pmatrix}
} \, .
\label{eq:Connection}
\end{equation}
We can then think of the transport equation~\eqref{eq:TransportNew} as
simply parallel transport under this new connection $\tilde{\cd}$ in
the direct-sum bundle,
\begin{equation}
\label{eq:cd-tilde-is-zero}
  \tilde{\cd}_{a} X^{B} = 0\,.
\end{equation}
In this bundle viewpoint, the questions related to the existence of
solutions to the transport equations~\eqref{eq:TransportNew} can be
cast in terms of conditions on the curvature of the connection.
We thus compute this curvature in the next part, and we give a
necessary condition for the existence of solutions in the part
thereafter.

\subsection{Curvature of the connection and holonomy of the transport equation}
\label{subsec:Curvature}

Any connection $\mathcal{D}$ on a vector bundle has an associated
curvature tensor, by virtue of the linearity of the map
\begin{equation}
  (\mathcal{D}_{u}\mathcal{D}_{v} - \mathcal{D}_{v}\mathcal{D}_{u} -
  \mathcal{D}_{[u,v]})X = R^{(\mathcal{D})}(u,v)X\,,
\end{equation}
where $u,v$ are two arbitrary tangent vectors on the base manifold
(they are {\it not} related to the vectors $u^a$ and $v^a$ of
Sec.~\ref{subsec:MomPrescrip}), and $R^{(\mathcal{D})}(-,-)$
is a two-form taking values in linear transformations on the fiber
space.
If we work in indices and in a holonomic frame then we can write
\begin{equation}
  (\mathcal{D}_a\mathcal{D}_b - \mathcal{D}_b\mathcal{D}_a)X^{C} =
  R^{(\mathcal{D})}_{ab}{}^{C}{}_{E}X^{E}\,.
\end{equation}

Let us start by presenting the curvature of the connection $\cd$ on
$\mathcal{B}$.  This can be derived by working with the basic
connection $\cd$ on $\TM$ that acts diagonally on the two summands of
$\mathcal{B}$.  As a shorthand we will simply write
$R_{ab}{}^{C}{}_{E}$ in place of $R^{(\cd)}_{ab}{}^{C}{}_{E}$.
In our index notation, we find
\begin{equation}
\label{eq:R-cd-on-B}
  R_{ab}{}^{C}{}_{E} =
\begin{matrix}
  \mbox{\tiny{C=c}} \\ \mbox{\tiny{C=[cd]}}
\end{matrix}
\overset{
\begin{matrix}
  \makebox[\widthof{$R_{ab}{}^{c}{}_{e}$}]{\tiny{E=e}}
& \makebox[\widthof{$2R_{ab\phantom{[c}[e}^{\phantom{ab}[c}\delta^{d]}_{f]}$}]{\tiny{E=[ef]}}
\end{matrix}
}{
\begin{pmatrix}
R_{ab}{}^{c}{}_{e} & 0 \\
0 & 2R_{ab\phantom{[c}[e}^{\phantom{ab}[c}\delta^{d]}_{f]}
\end{pmatrix}
} \, .
\end{equation}

Now we may compute the curvature, $\tilde{R}$, of our new connection,
$\tilde{\cd}$.  We will make use of the difference between the two
connections, $\tilde{\cd}_{a}X^{B} =
\cd_{a}X^{B}+\Gamma_{a}{}^{B}{}_{D}X^{D}$.  From the connection
coefficients, we can find $\tilde{R}$ in terms of $R$.  A
straightforward calculation gives (again in a holonomic frame),
\begin{multline}
\label{eq:R-tilde-in-terms-of-R}
  \tilde{R}_{ab}{}^{C}{}_{E} = R_{ab}{}^{C}{}_{E}
  + \cd_{a}\Gamma_{b}{}^{C}{}_{E}- \cd_{b}\Gamma_{a}{}^{C}{}_{E}\\
  +\Gamma_{a}{}^{C}{}_{G}\Gamma_{b}{}^{G}{}_{E}
  -\Gamma_{b}{}^{C}{}_{G}\Gamma_{a}{}^{G}{}_{E}\,.
\end{multline}
Now we will combine Eq.~\eqref{eq:Connection} and
Eq.~\eqref{eq:R-cd-on-B} to compute the coefficients of $\tilde{R}$ in
indices.  First, ``squaring'' the matrix in Eq.~\eqref{eq:Connection}
gives
\begin{equation}
  \label{eq:Gamma-squared}
  \Gamma_{a}{}^{C}{}_{G}\Gamma_{b}{}^{G}{}_{E} =
  \begin{matrix}
    \mbox{\tiny{C=c}} \\ \mbox{\tiny{C=[cd]}}
\end{matrix}
\overset{
\begin{matrix}
  \makebox[\widthof{$2\kappa R^{c}{}_{abe}$}]{\tiny{E=e}}
& \makebox[\widthof{$2\kappa \delta_{\phantom{[}a}^{[c}R^{d]}{}_{bef}$}]{\tiny{E=[ef]}}
\end{matrix}
}{
\begin{pmatrix}
2\kappa R^{c}{}_{abe}& 0 \\
0 & 2\kappa \delta_{\phantom{[}a}^{[c}R^{d]}_{\phantom{d]}bef}
\end{pmatrix}
} \, .
\end{equation}
Combining Eq.~\eqref{eq:R-cd-on-B}, Eq.~\eqref{eq:Gamma-squared},
taking the gradient of Eq.~\eqref{eq:Connection}, antisymmetrizing,
and using Bianchi identities, we find
\begin{align}
  \label{eq:R-tilde}
&  \tilde{R}_{ab}{}^{C}{}_{E}=  \nonumber \\
&  \begin{matrix}
    \mbox{\tiny{C=c}} \\ \mbox{\tiny{C=[cd]}}
\end{matrix}
\overset{
\begin{matrix}
  \makebox[\widthof{$(1-2\kappa) R_{ab}{}^{c}{}_{e}$}]{\tiny{E=e}}
& \makebox[\widthof{$2R_{ab}{}^{[c}{}_{[e}\delta^{d]}_{f]} + 4\kappa \delta_{[a}^{[c}R^{d]}_{\phantom{d]}b]ef}$}]{\tiny{E=[ef]}}
\end{matrix}
}{
\begin{pmatrix}
(1-2\kappa) R_{ab}{}^{c}{}_{e}& \kappa \cd^{c} R_{abef} \\
0 & 2R_{ab \phantom{[c}[e}^{\phantom{ab}[c} \delta^{d]}_{f]} + 4\kappa \delta_{[a}^{[c}R^{d]}_{\phantom{d]}b]ef}
\end{pmatrix}
} \, .
\end{align}

\figholonomy

From this calculation, we can directly read off the holonomy that the
vector $X^{E}=(P^{e},J^{[ef]})$ acquires when transported (via the new
connection $\tilde{\cd}$) around a ``coordinate rectangle.''  We
construct such a rectangle based at point $\mathcal{P}$, first
going in direction $u^{a}$, then $v^{b}$, then back along $-u^{a}$,
and finally back along $-v^{b}$ (see Fig.~\ref{fig:holonomy}).
Starting with vector $X^{E}$, after the
transport we will have the new vector
\begin{equation}
  X^{E} \xrightarrow{\circlearrowleft_{uv}} X^{E} + \delta X^{E} \, ,
\end{equation}
where the deviation vector is given by
\begin{equation}
  \label{eq:delta-X-expr}
  \delta X^{C} = -u^{a}v^{b}\tilde{R}_{ab}{}^{C}{}_{E} X^{E}\,.
\end{equation}
Combining Eqs.~\eqref{eq:R-tilde} and \eqref{eq:delta-X-expr}, we find
\begin{equation}
\begin{pmatrix}
  \delta P^{c} \\ \delta J^{cd}
\end{pmatrix}
=
\begin{pmatrix}
  - u^{a}v^{b} \left[ (1-2\kappa)R_{ab}{}^{c}{}_{e} P^{e} +
    \kappa \cd^{c} R_{abef}J^{ef}
  \right] \\
  - u^{a}v^{b} \left(
    2R_{ab\phantom{[c}[e}^{\phantom{ab}[c}\delta^{d]}_{f]} + 4\kappa \delta_{[a}^{[c}R^{d]}_{\phantom{d]}b]ef}
  \right) J^{ef}
\end{pmatrix}
\,.
\end{equation}
Using the geometric bitensor approach in~\cite{Vines:2014uoa} reproduces
this more algebraic calculation~\cite{VinesNicholsInPrep}.

\subsection{Existence of solutions in an extended region}
\label{sec:exist-solut-an}

From the calculation of $\tilde{R}_{ab}{}^{C}{}_{E}$, we can now state
a necessary condition for the existence of solutions to
Eq.~\eqref{eq:cd-tilde-is-zero} in an extended region.  Suppose a
solution $\mathcal{X}^{C}$ exists in an extended region that includes the
point $\mathcal{P}$, and it takes the value $\mathcal{X}^{C}(\mathcal{P})$.
Then, a necessary condition for the solution's existence is that under
transport about an arbitrary coordinate rectangle determined by
$u^{a}, v^{b}$, the value $\mathcal{X}^{C}(\mathcal{P})$ returns to itself
(i.e., there is a vanishing deviation vector $\delta \mathcal{X}^{C}=0$).
From Eq.~\eqref{eq:delta-X-expr}, we have
\begin{align}
  u^{a}v^{b}&\tilde{R}_{ab}{}^{C}{}_{E} \mathcal{X}^{E} = 0 \quad \forall u,v \\
\therefore\quad &  \tilde{R}_{ab}{}^{C}{}_{E} \mathcal{X}^{E} = 0 \,.
\end{align}
To interpret this condition, let us treat $\tilde{R}_{ab}{}^{C}{}_{E}$
as a linear map $\tilde{R}: \mathcal{B} \to
\Lambda^{2}\TsM \times \mathcal{B}$.  Then, this necessary
condition is that the map $\tilde{R}$ has a nontrivial kernel.

We have checked through an explicit coordinate-component calculation
in the Kerr spacetime, in Boyer-Lindquist coordinates, that
$\tilde{R}$ with $\kappa=-1/4$ has a one-dimensional kernel.
Thus, the space of solutions to
Eq.~\eqref{eq:cd-tilde-is-zero} is a one-dimensional linear space.
This is equivalent to the timelike Killing field and dual KY tensor
of Sec.~\ref{sec:Transport} being unique solutions
to~\eqref{eq:TransportNew} up to an overall multiplicative constant.

It is important to note, though, that this condition is only a
necessary condition, and not a \emph{sufficient} condition, for the
existence of solutions in an extended region.  It is easy to see why
this is true by looking at the case $\kappa=1/2$ in
Eq.~\eqref{eq:R-tilde}.  We see that the 4-dimensional subspace
$T\mathcal{M} \subset \mathcal{B}$ is automatically within the kernel
of $\tilde{R}$ for $\kappa=1/2$.  In fact, we have verified in Kerr
that this is the entirety of the kernel.  However, the
system~\eqref{eq:cd-tilde-is-zero} with $\kappa=1/2$ does not have
solutions
in an extended region.  If one starts at a point $\mathcal{P}$ with
data $\mathcal{X}^{C}(\mathcal{P})=(P^{c},0)$ with $P^{c}\neq 0$, then at some
nearby point $Q\neq \mathcal{P}$, the transported data will have
rotated out of the kernel, such that $J^{[cd]}(Q)\neq 0$.

A stronger condition is required for sufficiency.  This condition
comes from Frobenius' theorem for a tangent distribution to be
integrable into a submanifold (i.e., for the distribution to be
involutive).  We leave an investigation of this sufficient
condition for future work.

\section{Conclusions}
\label{sec:Conclusions}

In this paper, we have introduced a method for observers to measure
a kind of linear and angular momentum at a spacetime point from the
Riemann tensor and its derivatives, and we also proposed
a method to transport these momenta.
These measurement and transport procedures are in the same spirit as
those of Paper~I, but they also contain some important refinements.
The refinements are designed so that
observers who use both these procedures in stationary, vacuum
regions of asymptotically flat spacetimes will
find that their measurements are consistent with one another,
asymptotically as $r \to \infty$.  Thus, the procedures give a simple
operational meaning to the linear and angular momentum of the
spacetime in stationary regions of $\scri$.

In this paper, the transport and measurement procedures are much
more closely coupled to one another than they were in Paper~I.
We introduced a one-parameter family of transport equations, and we
found that for a unique value of this parameter ($\kappa=-1/4$),
there is a seven-parameter family of approximate solutions that can be
transported independently of path in the $r \to \infty$ limit for
stationary, vacuum spacetimes.
Our measurement procedure was designed to reproduce one of these
solutions in the appropriate limit.

We also explained how the transport equation for linear and angular
momentum could be understood as parallel transport for a specific
connection on a certain direct-sum vector bundle.
We computed the curvature of this connection and used it to find
the holonomy for an infinitesimal quadrilateral loop.
From the curvature, we could also formulate a necessary (though
not sufficient) condition for the existence of global sections
of the bundle.

A similar procedure is not possible, we conjecture, at higher order in
powers of $1/r$, in a general stationary, asymptotically-flat, vacuum
spacetime.  That is, the order to which we have worked in this
article is the highest possible order where a procedure of
path-independent transport is possible for $(P^{a},J^{ab})$.  This is
because to the present order, all stationary, asymptotically-flat,
vacuum spacetimes can be matched with an expansion of the Kerr
spacetime.  However, if we were to specialize to the Kerr spacetime
and to continue to expand to higher orders in $1/r$, we conjecture that
the transport equations~\eqref{eq:FieldEqs} would constrain
$J^{\mu\nu}_{(1)}, J^{\mu\nu}_{(2)},\ldots$
to have the form of the dual to the Killing-Yano tensor, expanded to
the appropriate order.
We leave investigation of this conjecture to future work.

Other possible future directions include an
exploration of sufficient conditions for global sections to
exist using Frobenius' theorem and an exploration of
BMS-type ambiguities in angular
momentum in intermittently stationary spacetimes by the
measurement and transport procedures developed here.

\acknowledgments

L.C.S.~acknowledges that support for this work was provided by NASA
through Einstein Postdoctoral Fellowship Award Number PF2--130101
issued by the Chandra X-ray Observatory Center, which is operated by
the Smithsonian Astrophysical Observatory for and on behalf of the
NASA under contract NAS8--03060, and further acknowledges support from
the NSF grant PHY--1404569.
E.F.~and D.N.~acknowledge support of NSF grants PHY--1404105 and
PHY--1068541.

\bibliography{Refs}

\end{document}